\documentclass[reprint,prl]{revtex4-1}
\usepackage[pdftex]{graphicx}
\usepackage[]{color}

\definecolor{comment}{rgb}{0.16,0.54,.16}

\begin{document}

\title{Condensed phase effects on the electronic momentum distribution in the warm dense matter regime }
\author{Brian A. Mattern}
\author{Gerald T. Seidler}
\email{seidler@uw.edu}
\author{Joshua J. Kas}
\affiliation{Department of Physics, University of Washington, Seattle, WA 98195-1560}
\date{\today}

\begin{abstract}

  We report \textit{ab initio} calculations of the valence electron momentum
  distribution function $n(p)$ and dynamic structure factor for warm dense Be
  at Mbar pressures.  We observe an unexpected, strong reshaping of the
  Compton profile upon increasing density, even well before any significant
  core-wavefunction overlap or electride behavior occurs.  We propose that
  this nonperturbative effect, which is due to a growing influence on $n(p)$
  of the orthogonalization of valence and core electron wavefunctions with
  increasing density, is observable by inelastic x-ray scattering at x-ray
  free-electron lasers and large-scale laser-shock heating facilities, and may
  also be more generally important for thermodynamic properties of dense,
  partially-ionized plasmas. 

  (submitted Phys. Rev. Lett. Aug. 2013)

\end{abstract}

\maketitle

There is growing interest in dense states of matter intermediate between
traditional condensed phase systems and fully-ionized dense plasmas. These
``warm dense matter'' (WDM) states present unique scientific opportunities
with special relevance to planetary and stellar
conditions\cite{Huser2005,Amadou2013,Remington2006,Wilson2012b}, with obvious
importance for the early stages of compression in inertial confinement fusion
(ICF)\cite{Lindl2004}, and with additional broad, long-term scientific
potential when considered as the next-stage of evolution of the available
thermodynamic parameter space for the condensed phase community.

As with any new material regime, one must determine the dominant microphysics
that establishes the resulting (macroscopic) thermodynamic and statistical
properties.  For WDM, this  is complicated by the lack of simplifying features
present in either the cold, ordered limit of condensed matter or the hot,
non-degenerate limit of fully-ionized dilute plasmas; unsurprisingly, at
present, no broadly-applicable, first-principles treatment of WDM is available.
One must instead combine methods from these opposing
limits\cite{Glenzer2009,Redmer2010,Sjostrom2012,Karasiev2012,Karasiev2012b} and then seek
comparison with the limited, but growing, body of WDM experimental results.

As a case in point, a common theoretical approach to
WDM\cite{Gregori2003,Gregori2004,Glenzer2009} treats free or ionized electrons
in the random phase approximation (RPA)\cite{BohmPines1953} with perturbative
corrections for the electron-ion scattering via the Born-Mermin approximation
(BMA)\cite{Selchow2001,Mermin1970}. This approach necessarily assumes a weak
electron-ion interaction and has seen extensive use in the interpretation of
inelastic x-ray scattering from
WDM\cite{Lee2009,Kritcher2011,Fortmann2012,Ma2013}.

By contrast, in crystalline materials and other condensed-phase systems it is
the electron-ion interaction, both through the Coulombic potential and
orthogonalization between core and valence electrons, that plays the dominant
role in the overall electronic
structure\cite{Bloch1929,Ashcroft,Ashcroft1966}. In addition, at elevated
densities, where core wavefunctions begin to overlap, an interesting
combination of free-energy effects constrained by valence-core
orthogonalization (VCO) can lead to structural changes and novel phases known as
electrides in which the valence electrons re-localize in the interstitial
spaces between atomic sites\cite{Neaton1999,Neaton2001,Rousseau2008}. This
behavior at low temperatures hints at the possibility of strong influence of
the electron-ion interaction, and in particular VCO, on the electronic
structure of WDM.  This conclusion, which runs contrary to the commonly stated
perspective that the electron-ion interaction becomes progressively less
important at high plasma densities\cite{LandauStatMech}, is further supported
by the growing use in WDM molecular dynamics simulations of modern
density-functional theory methods, such as projector-augmented
wave calculations, that substantially include VCO
effects\cite{Plagemann2012,Vorberger2013}.

Here, we report a detailed theoretical study of the valence electron momentum
distribution $n(p)$ for warm dense Be.  In addition to being a candidate
ablator material for ICF, and thus undergoing extensive study in the WDM
context\cite{Lee2009,Kritcher2011,Fortmann2012}, Be has seen thorough
high-resolution synchrotron
study\cite{Itou1998,Huotari2000,Huotari2007,Huotari2010}.  Our results
illustrate that VCO, which has a long history in the context of electronic
structure calculations\cite{Herring1940,Blochl1994} and which plays an
important role in determining the ambient momentum distribution of Be and
other materials\cite{Pandey1973,Cooper1974,Rennert1981,Bellaiche1997},
steadily grows in importance upon increasing density.  This observation may
have significant consequences for calculation of the equation of state or
thermodynamic susceptibilities (e.g., compressibility) for WDM but here we
focus on an issue of central experimental importance: $n(p)$ is a primary
experimental observable, appearing as a direct contributor to the dynamic
structure factor measured by non-resonant inelastic x-ray scattering (NIXS) in
the noncollective scattering regime and subsequently used to infer temperature
and density of WDM\cite{Glenzer2009}.

\begin{figure*}
  \includegraphics[width=180mm]{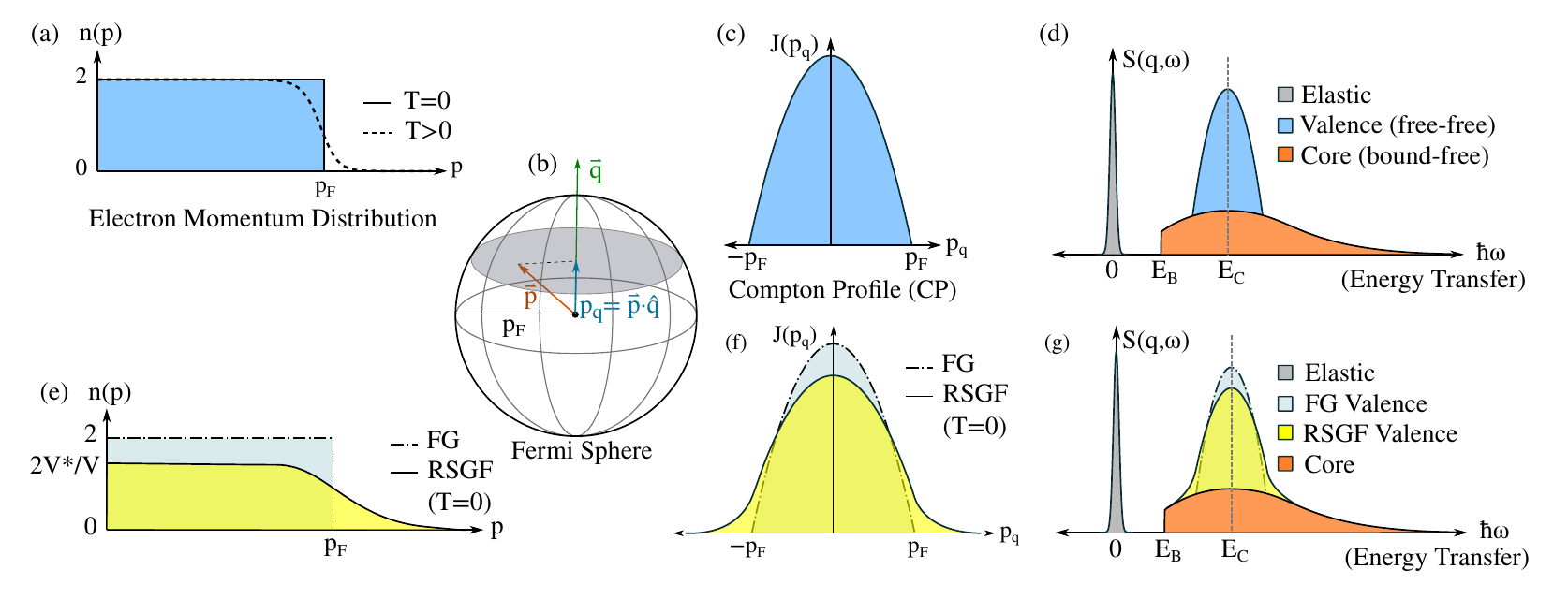}
  \caption{\
    Relationship between electronic momentum distribution and nonresonant IXS
    spectrum in the impulse approximation. (a)-(d): a non-interacting
    electron gas. (e)-(g) an electron gas with strong electron-ion
    interaction, such as due to valence-core orthogonalization (VCO). See the
    text for discussion. 
  }
  \label{fig:overview}
\end{figure*}

NIXS experiments measure the dynamic structure factor $S(q, \omega)$
which determines the relative probability of transferring momentum $\hbar q$
and energy $\hbar\omega$ from the probe radiation to the sample in the
scattering process\cite{Schuelke}.  At large $q$, the interpretation of NIXS
simplifies due to the impulse approximation (IA)\cite{Eisenberger1970} wherein
the potential before and after the scattering process cancel implying that
\begin{equation}
  \hbar\omega = \frac{\hbar^2 q^2}{2m} + \frac{\hbar\vec{q} \cdot \vec{p}}{m},
  \label{eq:doppler}
\end{equation}
where $\vec{p}$ is the scattering electron's initial momentum.
This Doppler broadened Compton scattering is entirely determined by the
electronic momentum distribution $n(p)$.

In Fig.~\ref{fig:overview} (a-d), we illustrate the relationship between
$n(p)$ and the NIXS spectrum using a simple Fermi gas model for the valence
electrons.  In panel (a), we show the Fermi radial momentum distribution. This
corresponds to the Fermi sphere shown in panel (b).  
In the shaded intersection between the Fermi sphere and a plane
perpendicular to $\vec{q}$, the Doppler shift, and thus energy transfer, is
identical.  The area of this intersection as a function of displacement from
the origin $p_q \equiv \vec{p}\cdot\hat{q}$ defines the Compton profile (CP)
$J(p_q)$, shown in panel (c). According to Eq.~(\ref{eq:doppler}), the CP is
displaced by the Compton shift $E_{\rm C} = \hbar^2q^2/2m_e$ and stretched by $\hbar q/m$ to obtain
$S(q,\omega)$.  In panel (d), we show this along with a representative
contribution from tightly-bound core electrons ($E_B$ is the core-state
binding energy)\cite{Mattern2013}.

Each of temperature, free-electron density, electron-electron interactions and
electron-ion interactions influences $n(p)$.  As shown in
Fig.~\ref{fig:overview}(e), interactions, if strong, have a global impact on
$n(p)$ and the shape of the CP by moving occupation from states below the
single-electron Fermi level to states above, even at $T=0$.  For an
interacting Fermi gas at the density of ambient Be, electron-electron
correlations result in a promotion of $5\%$ of electrons to states above the
Fermi level\cite{Supplemental,Kas2013}.  As density increases, this effect
diminishes.  On the other hand, we show here that electron-ion interactions,
which already have a stronger influence on electron occupation at ambient
conditions\cite{Huotari2000,Rennert1981} (with $20\%$ of electrons promoted
above $p_F$\cite{Supplemental}) have a growing, dramatic impact on $n(p)$ as
density increases.  This effect, and its consequences are shown schematically
in Fig.~\ref{fig:overview} (e)-(g) and are presented in detail in the
remainder of this letter.

\begin{figure}
  \begin{center}
    \includegraphics{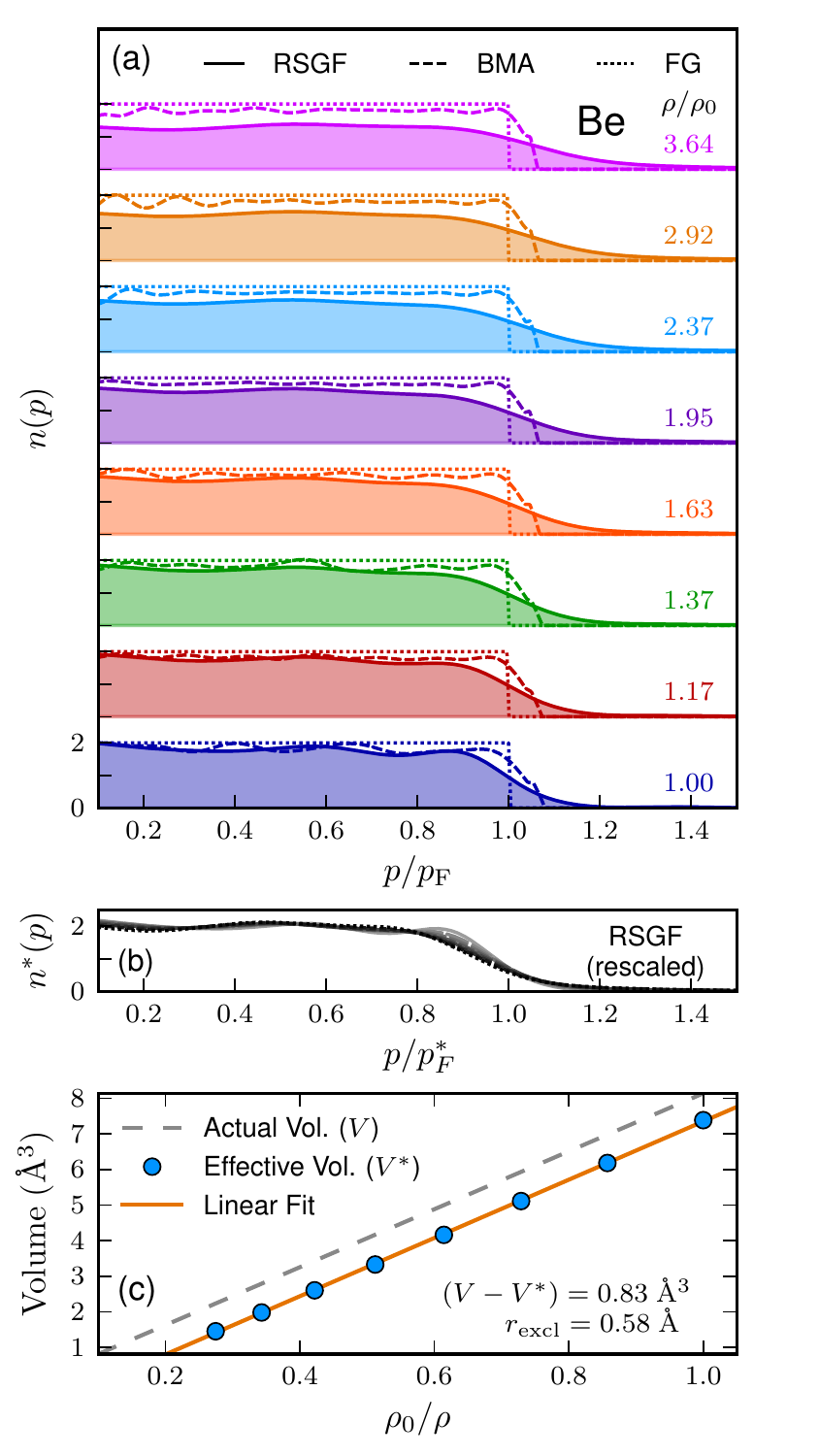}
  \end{center}
  \caption{Theoretical electronic momentum density of hcp Be as a function of
    atomic density $\rho$ in terms of ambient density $\rho_0$. The highest density shown
    corresponds to a pressure of $\sim 8$ Mbar\cite{Benedict2009}.
    (a) Comparison between calculations using the non-perturbative real-space
    Green's function method (RSGF), the perturbative Born-Mermin approximation
    (BMA) and the non-interacting Fermi gas (FG). The momentum $p$ is scaled by
    the Fermi momentum $p_{\rm F}$.
    (b) RSGF calculations rescaled by an effective volume determined by the
    average occupation at low momenta. Curves at higher density are drawn
    darker.
    (c) The effective volume per atom as a function of inverse compression.
  }
  \label{fig:rhop}
\end{figure}

In Fig.~\ref{fig:rhop}, we present calculations of the spherically-averaged
radial momentum distribution $n(p)$ for hcp beryllium at $T=0$ as a function of
atomic density using a real-space Green's function (RSGF)
method\cite{Mattern2012,Rehr2009,RehrAlbers2000} that treats the electron-ion
interaction non-perturbatively, including the effects of VCO\@.  This is
compared to calculations for a non-interacting Fermi gas (FG) and to
calculations that include electron-ion interactions perturbatively via the
BMA\cite{Selchow2001}.  The distribution function shown in Fig.~\ref{fig:rhop}
is the number of electrons per momentum eigenstate $n(p)$, which is related to
the momentum density $\rho(p)$ by $n(p) = ((2\pi)^3/V) \rho(p)$.  The abscissa
in Fig.~\ref{fig:rhop} has been rescaled by the Fermi momentum, $p_F$ to
capture the complete density-dependence of the Fermi gas n(p). The perturbative
electron-ion interaction in the BMA results in a slight decrease in occupation
at all momenta compensated for by increased occupation above $p_F$.  Evidently,
the bulk of the BMA density dependence is also captured by the rescaling by
$p_F$. The RSGF calculations, on the other hand, show a marked density
dependence differing from that of the free or weakly interacting FG\@.  We note
two primary features.  First, $n(p)$ is nearly uniformly decreased at low $p$
and this depletion gets stronger with increased density.  Second, the relative
weight of the tail above $p_F$ increases with density.

\begin{figure}
  \begin{center}
    \includegraphics{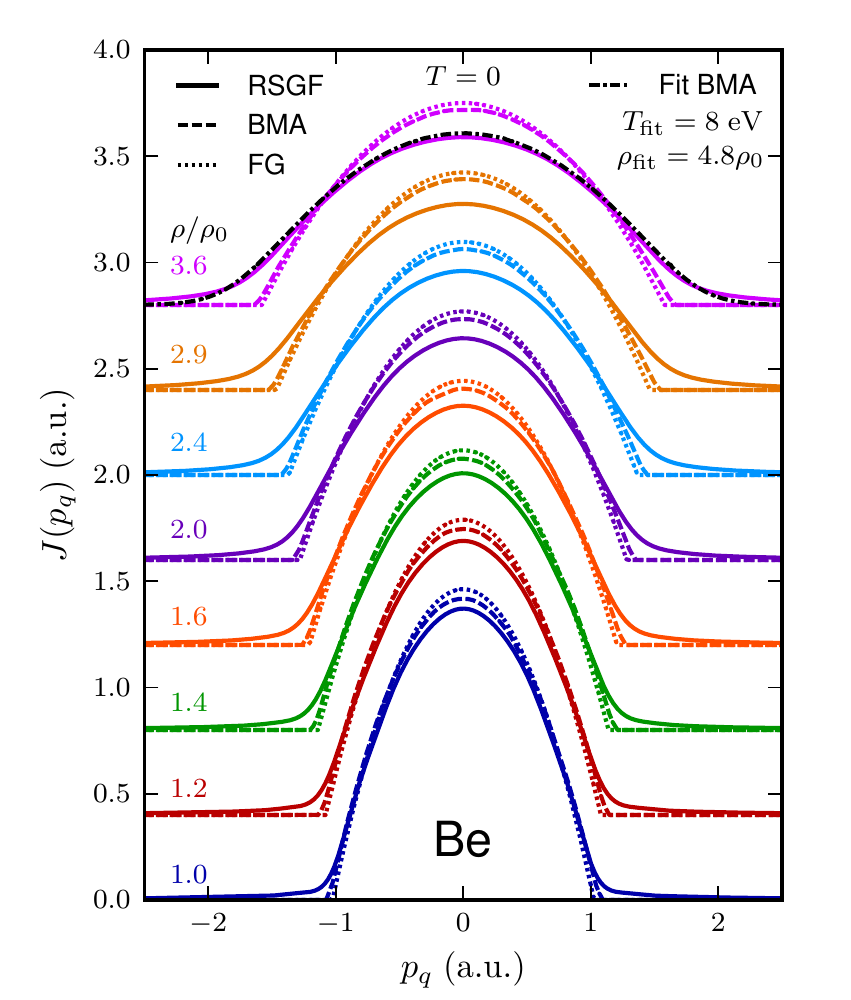}
  \end{center}
  \caption{Theoretical Compton profiles for Be metal as a function of density
    at T=0. The real-space Green's function (RSGF, solid curves) is compared to
    the Born-Mermin Approximation (BMA, dashed lines) and the Fermi
    gas (FG, dotted lines). In addition, for the highest density, we show the
    result of fitting a BMA calculation with adjustable temperature and density
    to the RSGF curve.
  }
  \label{fig:jpq}
\end{figure}

In Fig.~\ref{fig:jpq}, we show Compton profiles that correspond to the
momentum distributions from Fig.~\ref{fig:rhop}. The slight reshaping of the
RSGF CP relative to the FG at ambient density is in good agreement with
high-resolution experimental data\cite{Mattern2012}.  As density increases,
the differences between RSGF and FG become quite dramatic: the electron-ion
interaction results in a significant reshaping of the CP, shifting weight from
the peak out into the high-momentum tails.

The broad, nearly uniform depletion of valence occupation for $p < p_F$
indicates gross changes in the available phase space. Interestingly, these
effects have a simple relationship to atomic density that suggests an
overwhelming importance of VCO for $n(p)$ of solids or partially-ionized
plasmas at high densities.  In Fig.~\ref{fig:rhop}b we show a simple rescaling
of all RSGF $n(p)$ by the apparent effective free volume per atom
inferred from $\bar{n}$, the average occupancy of $n(p)$ at low $p$ (see
Fig.~\ref{fig:rhop}c), i.e., $V^* = \bar{n}V/2$.  The simple offset in the
effective volume $V^*$ indicates that the overall phenomenon can be discussed,
at least qualitatively, as an excluded volume effect encompassing the strong
constraint imposed by VCO on the valence electron wavefunctions in the
vicinity of the ion core.  Numerically, $V_{\rm excl}=0.83$ \AA$^3$, which is
the volume in which 98\% of the $1s$ electrons are contained.  There are clear
similarities between these results, where VCO with Hartree-Fock bound states
has been imposed, and the simpler hard-sphere model introduced by Rousseau and
Ashcroft as a heuristic tool for electride formation in Na\cite{Rousseau2008}.
In each case, the key physical insight is a substantial constraint on the
valence electron wavefunctions over a fixed volume per atom, independent of
overall atomic density.

\begin{figure*}[ht]
  \begin{center}
    \includegraphics{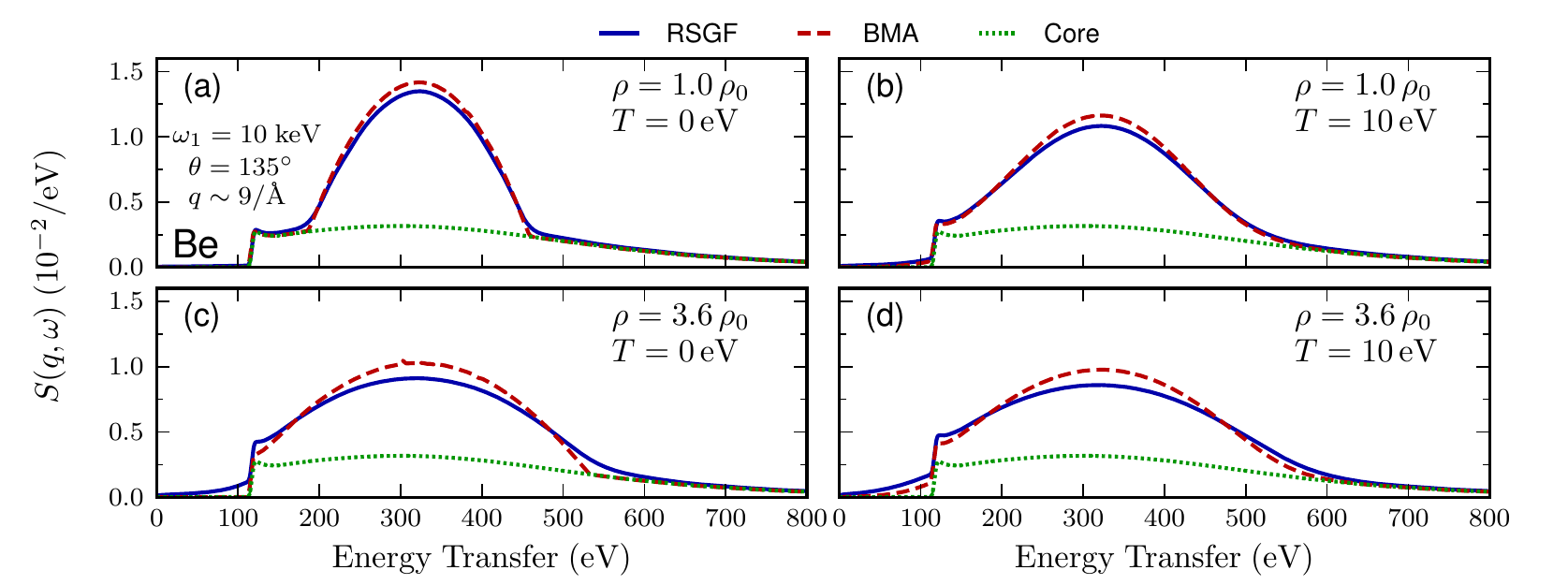}
  \end{center}
  \caption{Simulated NIXS spectra.
    (a) Ambient density and temperature
    (b) Isochoric heating
    (c) Adiabatic compression
    (d) Shock Compression/Heating.
  All curves are for a 10-keV incident energy at $135^\circ$ scattering angle
  and include 5-eV broadening. }
  \label{fig:comparison}
\end{figure*}

These results establish that calculations of the momentum density function for
WDM must necessarily include nonperturbative effects of the electron-ion
interaction at the level of wavefunctions or Green's function propagators even
well before electride formation is reached.  We now discuss the consequences
of this observation. First, a full understanding of the consequences of VCO on
the EOS or thermodynamic susceptibilities, especially the compressibility,
will require further work.  That being said, the strong non-free-electron-like
modification of $n(p)$, and thus also the kinetic energy\cite{Epstein1973},
with compression makes it clear that its consequences cannot, \textit{a
priori}, be ignored.

Second, moving to experimental consequences, we return to Fig.~\ref{fig:jpq}
where, for the highest-density calculations, we show a fit to the RSGF
predictions using the BMA with adjustable $\rho$ and $T$.  The large weight in
the RSGF tail is reproduced by the BMA model only when increasing the density
by 30\% and increasing the temperature from 0 to 8 eV.  While the exact
consequences for interpretation or reinterpretation of experiment will require
further work, the message is clear: failure to include the effects of VCO will
result in a systematic overestimate of either $T$, $n_e$, or both.

In Fig.~\ref{fig:comparison} we compare RSGF and BMA valence contributions to
synthetic NIXS spectra under the following four thermodynamic conditions:
(a)~ambient, (b)~isochoric heating, (c)~isothermal compression, and
(d)~shock-compression/heating.  High-temperature RSGF calculations were
performed by using a finite-temperature exchange-correlation
potential\cite{Perrot1984} and using the Fermi distribution for the occupation
of states both in the self-consistency loop and in the final calculation of
the Compton profile.\cite{MatternThesis} The ionic lattice was kept ordered.
All curves have been broadened by a 5-eV FWHM Gaussian.  At this resolution,
which is attainable using the seeded source at LCLS, the effects of VCO on the
Compton profile should be measurable.

One interesting future direction is to investigate the corresponding
momentum-space phenomenon associated with the real-space segregation of
valence charge density in proposed electride phases.  It may be the case that
electride formation can be more easily observed, especially in disordered
systems, by its influence on $n(p)$ rather than by its slight modification of
the static structure factor $S(q)$ in elastic scattering.

In conclusion, we have shown that non-perturbative effects of the electron-ion
interaction, such as core-valence orthogonalization, play an increasingly
important role in determining the momentum distribution of free electrons in
ordered dense matter as density is increased.  This will have a profound effect
on the interpretation of NIXS-based diagnostic measurements of WDM, and may
also be relevant for thermodynamic properties of dense, partially ionized
plasmas.

\begin{acknowledgments}
This work was supported by the US Department of Energy, Office of Science,
Fusion Energy Sciences and the National Nuclear Security Administration,
through grant DE-SC0008580.  We thank J. J. Rehr, S. Huotari and S. Hanson
for many helpful discussions.
\end{acknowledgments}


\begin{thebibliography}{50}%
\makeatletter
\providecommand \@ifxundefined [1]{%
 \@ifx{#1\undefined}
}%
\providecommand \@ifnum [1]{%
 \ifnum #1\expandafter \@firstoftwo
 \else \expandafter \@secondoftwo
 \fi
}%
\providecommand \@ifx [1]{%
 \ifx #1\expandafter \@firstoftwo
 \else \expandafter \@secondoftwo
 \fi
}%
\providecommand \natexlab [1]{#1}%
\providecommand \enquote  [1]{``#1''}%
\providecommand \bibnamefont  [1]{#1}%
\providecommand \bibfnamefont [1]{#1}%
\providecommand \citenamefont [1]{#1}%
\providecommand \href@noop [0]{\@secondoftwo}%
\providecommand \href [0]{\begingroup \@sanitize@url \@href}%
\providecommand \@href[1]{\@@startlink{#1}\@@href}%
\providecommand \@@href[1]{\endgroup#1\@@endlink}%
\providecommand \@sanitize@url [0]{\catcode `\\12\catcode `\$12\catcode
  `\&12\catcode `\#12\catcode `\^12\catcode `\_12\catcode `\%12\relax}%
\providecommand \@@startlink[1]{}%
\providecommand \@@endlink[0]{}%
\providecommand \url  [0]{\begingroup\@sanitize@url \@url }%
\providecommand \@url [1]{\endgroup\@href {#1}{\urlprefix }}%
\providecommand \urlprefix  [0]{URL }%
\providecommand \Eprint [0]{\href }%
\providecommand \doibase [0]{http://dx.doi.org/}%
\providecommand \selectlanguage [0]{\@gobble}%
\providecommand \bibinfo  [0]{\@secondoftwo}%
\providecommand \bibfield  [0]{\@secondoftwo}%
\providecommand \translation [1]{[#1]}%
\providecommand \BibitemOpen [0]{}%
\providecommand \bibitemStop [0]{}%
\providecommand \bibitemNoStop [0]{.\EOS\space}%
\providecommand \EOS [0]{\spacefactor3000\relax}%
\providecommand \BibitemShut  [1]{\csname bibitem#1\endcsname}%
\let\auto@bib@innerbib\@empty
\bibitem [{\citenamefont {Huser}\ \emph {et~al.}(2005)\citenamefont {Huser},
  \citenamefont {Koenig}, \citenamefont {Benuzzi-Mounaix}, \citenamefont
  {Henry}, \citenamefont {Vinci}, \citenamefont {Faral}, \citenamefont
  {Tomasini}, \citenamefont {Telaro},\ and\ \citenamefont
  {Batani}}]{Huser2005}%
  \BibitemOpen
  \bibfield  {author} {\bibinfo {author} {\bibfnamefont {G.}~\bibnamefont
  {Huser}}, \bibinfo {author} {\bibfnamefont {M.}~\bibnamefont {Koenig}},
  \bibinfo {author} {\bibfnamefont {A.}~\bibnamefont {Benuzzi-Mounaix}},
  \bibinfo {author} {\bibfnamefont {E.}~\bibnamefont {Henry}}, \bibinfo
  {author} {\bibfnamefont {T.}~\bibnamefont {Vinci}}, \bibinfo {author}
  {\bibfnamefont {B.}~\bibnamefont {Faral}}, \bibinfo {author} {\bibfnamefont
  {M.}~\bibnamefont {Tomasini}}, \bibinfo {author} {\bibfnamefont
  {B.}~\bibnamefont {Telaro}}, \ and\ \bibinfo {author} {\bibfnamefont
  {D.}~\bibnamefont {Batani}},\ }\href {\doibase 10.1063/1.1896375} {\bibfield
  {journal} {\bibinfo  {journal} {Physics of Plasmas}\ }\textbf {\bibinfo
  {volume} {12}},\ \bibinfo {eid} {060701} (\bibinfo {year}
  {2005})}\BibitemShut {NoStop}%
\bibitem [{\citenamefont {Amadou}\ \emph {et~al.}(2013)\citenamefont {Amadou},
  \citenamefont {Brambrink}, \citenamefont {Benuzzi-Mounaix}, \citenamefont
  {Huser}, \citenamefont {Guyot}, \citenamefont {Mazevet}, \citenamefont
  {Morard}, \citenamefont {de~Resseguier}, \citenamefont {Vinci}, \citenamefont
  {Myanishi}, \citenamefont {Ozaki}, \citenamefont {Kodama}, \citenamefont
  {Boehly}, \citenamefont {Henry}, \citenamefont {Raffestin},\ and\
  \citenamefont {Koenig}}]{Amadou2013}%
  \BibitemOpen
  \bibfield  {author} {\bibinfo {author} {\bibfnamefont {N.}~\bibnamefont
  {Amadou}}, \bibinfo {author} {\bibfnamefont {E.}~\bibnamefont {Brambrink}},
  \bibinfo {author} {\bibfnamefont {A.}~\bibnamefont {Benuzzi-Mounaix}},
  \bibinfo {author} {\bibfnamefont {G.}~\bibnamefont {Huser}}, \bibinfo
  {author} {\bibfnamefont {F.}~\bibnamefont {Guyot}}, \bibinfo {author}
  {\bibfnamefont {S.}~\bibnamefont {Mazevet}}, \bibinfo {author} {\bibfnamefont
  {G.}~\bibnamefont {Morard}}, \bibinfo {author} {\bibfnamefont
  {T.}~\bibnamefont {de~Resseguier}}, \bibinfo {author} {\bibfnamefont
  {T.}~\bibnamefont {Vinci}}, \bibinfo {author} {\bibfnamefont
  {K.}~\bibnamefont {Myanishi}}, \bibinfo {author} {\bibfnamefont
  {N.}~\bibnamefont {Ozaki}}, \bibinfo {author} {\bibfnamefont
  {R.}~\bibnamefont {Kodama}}, \bibinfo {author} {\bibfnamefont
  {T.}~\bibnamefont {Boehly}}, \bibinfo {author} {\bibfnamefont
  {O.}~\bibnamefont {Henry}}, \bibinfo {author} {\bibfnamefont
  {D.}~\bibnamefont {Raffestin}}, \ and\ \bibinfo {author} {\bibfnamefont
  {M.}~\bibnamefont {Koenig}},\ }\href {\doibase
  http://dx.doi.org/10.1016/j.hedp.2013.01.003} {\bibfield  {journal} {\bibinfo
   {journal} {High Energy Density Physics}\ }\textbf {\bibinfo {volume} {9}},\
  \bibinfo {pages} {243 } (\bibinfo {year} {2013})}\BibitemShut {NoStop}%
\bibitem [{\citenamefont {Remington}\ \emph {et~al.}(2006)\citenamefont
  {Remington}, \citenamefont {Drake},\ and\ \citenamefont
  {Ryutov}}]{Remington2006}%
  \BibitemOpen
  \bibfield  {author} {\bibinfo {author} {\bibfnamefont {B.~A.}\ \bibnamefont
  {Remington}}, \bibinfo {author} {\bibfnamefont {R.~P.}\ \bibnamefont
  {Drake}}, \ and\ \bibinfo {author} {\bibfnamefont {D.~D.}\ \bibnamefont
  {Ryutov}},\ }\href {\doibase 10.1103/RevModPhys.78.755} {\bibfield  {journal}
  {\bibinfo  {journal} {Reviews of Modern Physics}\ }\textbf {\bibinfo {volume}
  {78}},\ \bibinfo {pages} {755} (\bibinfo {year} {2006})}\BibitemShut
  {NoStop}%
\bibitem [{\citenamefont {Wilson}\ and\ \citenamefont
  {Militzer}(2012)}]{Wilson2012b}%
  \BibitemOpen
  \bibfield  {author} {\bibinfo {author} {\bibfnamefont {H.~F.}\ \bibnamefont
  {Wilson}}\ and\ \bibinfo {author} {\bibfnamefont {B.}~\bibnamefont
  {Militzer}},\ }\href {\doibase 10.1103/PhysRevLett.108.111101} {\bibfield
  {journal} {\bibinfo  {journal} {Phys. Rev. Lett.}\ }\textbf {\bibinfo
  {volume} {108}},\ \bibinfo {pages} {111101} (\bibinfo {year}
  {2012})}\BibitemShut {NoStop}%
\bibitem [{\citenamefont {Lindl}\ \emph {et~al.}(2004)\citenamefont {Lindl},
  \citenamefont {Amendt}, \citenamefont {Berger}, \citenamefont {Glendinning},
  \citenamefont {Glenzer}, \citenamefont {Haan}, \citenamefont {Kauffman},
  \citenamefont {Landen},\ and\ \citenamefont {Suter}}]{Lindl2004}%
  \BibitemOpen
  \bibfield  {author} {\bibinfo {author} {\bibfnamefont {J.~D.}\ \bibnamefont
  {Lindl}}, \bibinfo {author} {\bibfnamefont {P.}~\bibnamefont {Amendt}},
  \bibinfo {author} {\bibfnamefont {R.~L.}\ \bibnamefont {Berger}}, \bibinfo
  {author} {\bibfnamefont {S.~G.}\ \bibnamefont {Glendinning}}, \bibinfo
  {author} {\bibfnamefont {S.~H.}\ \bibnamefont {Glenzer}}, \bibinfo {author}
  {\bibfnamefont {S.~W.}\ \bibnamefont {Haan}}, \bibinfo {author}
  {\bibfnamefont {R.~L.}\ \bibnamefont {Kauffman}}, \bibinfo {author}
  {\bibfnamefont {O.~L.}\ \bibnamefont {Landen}}, \ and\ \bibinfo {author}
  {\bibfnamefont {L.~J.}\ \bibnamefont {Suter}},\ }\href {\doibase
  10.1063/1.1578638} {\bibfield  {journal} {\bibinfo  {journal} {Physics of
  Plasmas}\ }\textbf {\bibinfo {volume} {11}},\ \bibinfo {pages} {339}
  (\bibinfo {year} {2004})}\BibitemShut {NoStop}%
\bibitem [{\citenamefont {Glenzer}\ and\ \citenamefont
  {Redmer}(2009)}]{Glenzer2009}%
  \BibitemOpen
  \bibfield  {author} {\bibinfo {author} {\bibfnamefont {S.~H.}\ \bibnamefont
  {Glenzer}}\ and\ \bibinfo {author} {\bibfnamefont {R.}~\bibnamefont
  {Redmer}},\ }\href {\doibase 10.1103/RevModPhys.81.1625} {\bibfield
  {journal} {\bibinfo  {journal} {Reviews of Modern Physics}\ }\textbf
  {\bibinfo {volume} {81}},\ \bibinfo {pages} {1625} (\bibinfo {year}
  {2009})}\BibitemShut {NoStop}%
\bibitem [{\citenamefont {Redmer}\ and\ \citenamefont
  {R\"opke}(2010)}]{Redmer2010}%
  \BibitemOpen
  \bibfield  {author} {\bibinfo {author} {\bibfnamefont {R.}~\bibnamefont
  {Redmer}}\ and\ \bibinfo {author} {\bibfnamefont {G.}~\bibnamefont
  {R\"opke}},\ }\href {\doibase 10.1002/ctpp.201000079} {\bibfield  {journal}
  {\bibinfo  {journal} {Contributions to Plasma Physics}\ }\textbf {\bibinfo
  {volume} {50}},\ \bibinfo {pages} {970} (\bibinfo {year} {2010})}\BibitemShut
  {NoStop}%
\bibitem [{\citenamefont {Sjostrom}\ \emph {et~al.}(2012)\citenamefont
  {Sjostrom}, \citenamefont {Harris},\ and\ \citenamefont
  {Trickey}}]{Sjostrom2012}%
  \BibitemOpen
  \bibfield  {author} {\bibinfo {author} {\bibfnamefont {T.}~\bibnamefont
  {Sjostrom}}, \bibinfo {author} {\bibfnamefont {F.~E.}\ \bibnamefont
  {Harris}}, \ and\ \bibinfo {author} {\bibfnamefont {S.~B.}\ \bibnamefont
  {Trickey}},\ }\href {\doibase 10.1103/PhysRevB.85.045125} {\bibfield
  {journal} {\bibinfo  {journal} {Phys. Rev. B}\ }\textbf {\bibinfo {volume}
  {85}},\ \bibinfo {pages} {045125} (\bibinfo {year} {2012})}\BibitemShut
  {NoStop}%
\bibitem [{\citenamefont {Karasiev}\ \emph
  {et~al.}(2012{\natexlab{a}})\citenamefont {Karasiev}, \citenamefont
  {Sjostrom},\ and\ \citenamefont {Trickey}}]{Karasiev2012}%
  \BibitemOpen
  \bibfield  {author} {\bibinfo {author} {\bibfnamefont {V.~V.}\ \bibnamefont
  {Karasiev}}, \bibinfo {author} {\bibfnamefont {T.}~\bibnamefont {Sjostrom}},
  \ and\ \bibinfo {author} {\bibfnamefont {S.~B.}\ \bibnamefont {Trickey}},\
  }\href {\doibase 10.1103/PhysRevE.86.056704} {\bibfield  {journal} {\bibinfo
  {journal} {Phys. Rev. E}\ }\textbf {\bibinfo {volume} {86}},\ \bibinfo
  {pages} {056704} (\bibinfo {year} {2012}{\natexlab{a}})}\BibitemShut
  {NoStop}%
\bibitem [{\citenamefont {Karasiev}\ \emph
  {et~al.}(2012{\natexlab{b}})\citenamefont {Karasiev}, \citenamefont
  {Sjostrom},\ and\ \citenamefont {Trickey}}]{Karasiev2012b}%
  \BibitemOpen
  \bibfield  {author} {\bibinfo {author} {\bibfnamefont {V.~V.}\ \bibnamefont
  {Karasiev}}, \bibinfo {author} {\bibfnamefont {T.}~\bibnamefont {Sjostrom}},
  \ and\ \bibinfo {author} {\bibfnamefont {S.~B.}\ \bibnamefont {Trickey}},\
  }\href {\doibase 10.1103/PhysRevB.86.115101} {\bibfield  {journal} {\bibinfo
  {journal} {Phys. Rev. B}\ }\textbf {\bibinfo {volume} {86}},\ \bibinfo
  {pages} {115101} (\bibinfo {year} {2012}{\natexlab{b}})}\BibitemShut
  {NoStop}%
\bibitem [{\citenamefont {Gregori}\ \emph {et~al.}(2003)\citenamefont
  {Gregori}, \citenamefont {Glenzer}, \citenamefont {Rozmus}, \citenamefont
  {Lee},\ and\ \citenamefont {Landen}}]{Gregori2003}%
  \BibitemOpen
  \bibfield  {author} {\bibinfo {author} {\bibfnamefont {G.}~\bibnamefont
  {Gregori}}, \bibinfo {author} {\bibfnamefont {S.~H.}\ \bibnamefont
  {Glenzer}}, \bibinfo {author} {\bibfnamefont {W.}~\bibnamefont {Rozmus}},
  \bibinfo {author} {\bibfnamefont {R.~W.}\ \bibnamefont {Lee}}, \ and\
  \bibinfo {author} {\bibfnamefont {O.~L.}\ \bibnamefont {Landen}},\ }\href
  {\doibase 10.1103/PhysRevE.67.026412} {\bibfield  {journal} {\bibinfo
  {journal} {Phys. Rev. E}\ }\textbf {\bibinfo {volume} {67}},\ \bibinfo
  {pages} {026412} (\bibinfo {year} {2003})}\BibitemShut {NoStop}%
\bibitem [{\citenamefont {Gregori}\ \emph {et~al.}(2004)\citenamefont
  {Gregori}, \citenamefont {Glenzer}, \citenamefont {Rogers}, \citenamefont
  {Pollaine}, \citenamefont {Landen}, \citenamefont {Blancard}, \citenamefont
  {Faussurier}, \citenamefont {Renaudin}, \citenamefont {Kuhlbrodt},\ and\
  \citenamefont {Redmer}}]{Gregori2004}%
  \BibitemOpen
  \bibfield  {author} {\bibinfo {author} {\bibfnamefont {G.}~\bibnamefont
  {Gregori}}, \bibinfo {author} {\bibfnamefont {S.~H.}\ \bibnamefont
  {Glenzer}}, \bibinfo {author} {\bibfnamefont {F.~J.}\ \bibnamefont {Rogers}},
  \bibinfo {author} {\bibfnamefont {S.~M.}\ \bibnamefont {Pollaine}}, \bibinfo
  {author} {\bibfnamefont {O.~L.}\ \bibnamefont {Landen}}, \bibinfo {author}
  {\bibfnamefont {C.}~\bibnamefont {Blancard}}, \bibinfo {author}
  {\bibfnamefont {G.}~\bibnamefont {Faussurier}}, \bibinfo {author}
  {\bibfnamefont {P.}~\bibnamefont {Renaudin}}, \bibinfo {author}
  {\bibfnamefont {S.}~\bibnamefont {Kuhlbrodt}}, \ and\ \bibinfo {author}
  {\bibfnamefont {R.}~\bibnamefont {Redmer}},\ }\href {\doibase
  10.1063/1.1689664} {\bibfield  {journal} {\bibinfo  {journal} {Physics of
  Plasmas}\ }\textbf {\bibinfo {volume} {11}},\ \bibinfo {pages} {2754}
  (\bibinfo {year} {2004})}\BibitemShut {NoStop}%
\bibitem [{\citenamefont {Bohm}\ and\ \citenamefont
  {Pines}(1953)}]{BohmPines1953}%
  \BibitemOpen
  \bibfield  {author} {\bibinfo {author} {\bibfnamefont {D.}~\bibnamefont
  {Bohm}}\ and\ \bibinfo {author} {\bibfnamefont {D.}~\bibnamefont {Pines}},\
  }\href {\doibase 10.1103/PhysRev.92.609} {\bibfield  {journal} {\bibinfo
  {journal} {Phys. Rev.}\ }\textbf {\bibinfo {volume} {92}},\ \bibinfo {pages}
  {609} (\bibinfo {year} {1953})}\BibitemShut {NoStop}%
\bibitem [{\citenamefont {Selchow}\ \emph {et~al.}(2001)\citenamefont
  {Selchow}, \citenamefont {R\"opke}, \citenamefont {Wierling}, \citenamefont
  {Reinholz}, \citenamefont {Pschiwul},\ and\ \citenamefont
  {Zwicknagel}}]{Selchow2001}%
  \BibitemOpen
  \bibfield  {author} {\bibinfo {author} {\bibfnamefont {A.}~\bibnamefont
  {Selchow}}, \bibinfo {author} {\bibfnamefont {G.}~\bibnamefont {R\"opke}},
  \bibinfo {author} {\bibfnamefont {A.}~\bibnamefont {Wierling}}, \bibinfo
  {author} {\bibfnamefont {H.}~\bibnamefont {Reinholz}}, \bibinfo {author}
  {\bibfnamefont {T.}~\bibnamefont {Pschiwul}}, \ and\ \bibinfo {author}
  {\bibfnamefont {G.}~\bibnamefont {Zwicknagel}},\ }\href {\doibase
  10.1103/PhysRevE.64.056410} {\bibfield  {journal} {\bibinfo  {journal} {Phys.
  Rev. E}\ }\textbf {\bibinfo {volume} {64}},\ \bibinfo {pages} {056410}
  (\bibinfo {year} {2001})}\BibitemShut {NoStop}%
\bibitem [{\citenamefont {Mermin}(1970)}]{Mermin1970}%
  \BibitemOpen
  \bibfield  {author} {\bibinfo {author} {\bibfnamefont {N.~D.}\ \bibnamefont
  {Mermin}},\ }\href {\doibase 10.1103/PhysRevB.1.2362} {\bibfield  {journal}
  {\bibinfo  {journal} {Phys. Rev. B}\ }\textbf {\bibinfo {volume} {1}},\
  \bibinfo {pages} {2362} (\bibinfo {year} {1970})}\BibitemShut {NoStop}%
\bibitem [{\citenamefont {Lee}\ \emph {et~al.}(2009)\citenamefont {Lee},
  \citenamefont {Neumayer}, \citenamefont {Castor}, \citenamefont {D\"oppner},
  \citenamefont {Falcone}, \citenamefont {Fortmann}, \citenamefont {Hammel},
  \citenamefont {Kritcher}, \citenamefont {Landen}, \citenamefont {Lee},
  \citenamefont {Meyerhofer}, \citenamefont {Munro}, \citenamefont {Redmer},
  \citenamefont {Regan}, \citenamefont {Weber},\ and\ \citenamefont
  {Glenzer}}]{Lee2009}%
  \BibitemOpen
  \bibfield  {author} {\bibinfo {author} {\bibfnamefont {H.~J.}\ \bibnamefont
  {Lee}}, \bibinfo {author} {\bibfnamefont {P.}~\bibnamefont {Neumayer}},
  \bibinfo {author} {\bibfnamefont {J.}~\bibnamefont {Castor}}, \bibinfo
  {author} {\bibfnamefont {T.}~\bibnamefont {D\"oppner}}, \bibinfo {author}
  {\bibfnamefont {R.~W.}\ \bibnamefont {Falcone}}, \bibinfo {author}
  {\bibfnamefont {C.}~\bibnamefont {Fortmann}}, \bibinfo {author}
  {\bibfnamefont {B.~A.}\ \bibnamefont {Hammel}}, \bibinfo {author}
  {\bibfnamefont {A.~L.}\ \bibnamefont {Kritcher}}, \bibinfo {author}
  {\bibfnamefont {O.~L.}\ \bibnamefont {Landen}}, \bibinfo {author}
  {\bibfnamefont {R.~W.}\ \bibnamefont {Lee}}, \bibinfo {author} {\bibfnamefont
  {D.~D.}\ \bibnamefont {Meyerhofer}}, \bibinfo {author} {\bibfnamefont
  {D.~H.}\ \bibnamefont {Munro}}, \bibinfo {author} {\bibfnamefont
  {R.}~\bibnamefont {Redmer}}, \bibinfo {author} {\bibfnamefont {S.~P.}\
  \bibnamefont {Regan}}, \bibinfo {author} {\bibfnamefont {S.}~\bibnamefont
  {Weber}}, \ and\ \bibinfo {author} {\bibfnamefont {S.~H.}\ \bibnamefont
  {Glenzer}},\ }\href {\doibase 10.1103/PhysRevLett.102.115001} {\bibfield
  {journal} {\bibinfo  {journal} {Phys. Rev. Lett.}\ }\textbf {\bibinfo
  {volume} {102}},\ \bibinfo {pages} {115001} (\bibinfo {year}
  {2009})}\BibitemShut {NoStop}%
\bibitem [{\citenamefont {Kritcher}\ \emph {et~al.}(2011)\citenamefont
  {Kritcher}, \citenamefont {D\"oppner}, \citenamefont {Fortmann},
  \citenamefont {Ma}, \citenamefont {Landen}, \citenamefont {Wallace},\ and\
  \citenamefont {Glenzer}}]{Kritcher2011}%
  \BibitemOpen
  \bibfield  {author} {\bibinfo {author} {\bibfnamefont {A.~L.}\ \bibnamefont
  {Kritcher}}, \bibinfo {author} {\bibfnamefont {T.}~\bibnamefont {D\"oppner}},
  \bibinfo {author} {\bibfnamefont {C.}~\bibnamefont {Fortmann}}, \bibinfo
  {author} {\bibfnamefont {T.}~\bibnamefont {Ma}}, \bibinfo {author}
  {\bibfnamefont {O.~L.}\ \bibnamefont {Landen}}, \bibinfo {author}
  {\bibfnamefont {R.}~\bibnamefont {Wallace}}, \ and\ \bibinfo {author}
  {\bibfnamefont {S.~H.}\ \bibnamefont {Glenzer}},\ }\href {\doibase
  10.1103/PhysRevLett.107.015002} {\bibfield  {journal} {\bibinfo  {journal}
  {Phys. Rev. Lett.}\ }\textbf {\bibinfo {volume} {107}},\ \bibinfo {pages}
  {015002} (\bibinfo {year} {2011})}\BibitemShut {NoStop}%
\bibitem [{\citenamefont {Fortmann}\ \emph {et~al.}(2012)\citenamefont
  {Fortmann}, \citenamefont {Lee}, \citenamefont {D\"oppner}, \citenamefont
  {Falcone}, \citenamefont {Kritcher}, \citenamefont {Landen},\ and\
  \citenamefont {Glenzer}}]{Fortmann2012}%
  \BibitemOpen
  \bibfield  {author} {\bibinfo {author} {\bibfnamefont {C.}~\bibnamefont
  {Fortmann}}, \bibinfo {author} {\bibfnamefont {H.~J.}\ \bibnamefont {Lee}},
  \bibinfo {author} {\bibfnamefont {T.}~\bibnamefont {D\"oppner}}, \bibinfo
  {author} {\bibfnamefont {R.~W.}\ \bibnamefont {Falcone}}, \bibinfo {author}
  {\bibfnamefont {A.~L.}\ \bibnamefont {Kritcher}}, \bibinfo {author}
  {\bibfnamefont {O.~L.}\ \bibnamefont {Landen}}, \ and\ \bibinfo {author}
  {\bibfnamefont {S.~H.}\ \bibnamefont {Glenzer}},\ }\href {\doibase
  10.1103/PhysRevLett.108.175006} {\bibfield  {journal} {\bibinfo  {journal}
  {Phys. Rev. Lett.}\ }\textbf {\bibinfo {volume} {108}},\ \bibinfo {pages}
  {175006} (\bibinfo {year} {2012})}\BibitemShut {NoStop}%
\bibitem [{\citenamefont {Ma}\ \emph {et~al.}(2013)\citenamefont {Ma},
  \citenamefont {D\"oppner}, \citenamefont {Falcone}, \citenamefont {Fletcher},
  \citenamefont {Fortmann}, \citenamefont {Gericke}, \citenamefont {Landen},
  \citenamefont {Lee}, \citenamefont {Pak}, \citenamefont {Vorberger},
  \citenamefont {W\"unsch},\ and\ \citenamefont {Glenzer}}]{Ma2013}%
  \BibitemOpen
  \bibfield  {author} {\bibinfo {author} {\bibfnamefont {T.}~\bibnamefont
  {Ma}}, \bibinfo {author} {\bibfnamefont {T.}~\bibnamefont {D\"oppner}},
  \bibinfo {author} {\bibfnamefont {R.~W.}\ \bibnamefont {Falcone}}, \bibinfo
  {author} {\bibfnamefont {L.}~\bibnamefont {Fletcher}}, \bibinfo {author}
  {\bibfnamefont {C.}~\bibnamefont {Fortmann}}, \bibinfo {author}
  {\bibfnamefont {D.~O.}\ \bibnamefont {Gericke}}, \bibinfo {author}
  {\bibfnamefont {O.~L.}\ \bibnamefont {Landen}}, \bibinfo {author}
  {\bibfnamefont {H.~J.}\ \bibnamefont {Lee}}, \bibinfo {author} {\bibfnamefont
  {A.}~\bibnamefont {Pak}}, \bibinfo {author} {\bibfnamefont {J.}~\bibnamefont
  {Vorberger}}, \bibinfo {author} {\bibfnamefont {K.}~\bibnamefont {W\"unsch}},
  \ and\ \bibinfo {author} {\bibfnamefont {S.~H.}\ \bibnamefont {Glenzer}},\
  }\href {\doibase 10.1103/PhysRevLett.110.065001} {\bibfield  {journal}
  {\bibinfo  {journal} {Phys. Rev. Lett.}\ }\textbf {\bibinfo {volume} {110}},\
  \bibinfo {pages} {065001} (\bibinfo {year} {2013})}\BibitemShut {NoStop}%
\bibitem [{\citenamefont {Bloch}(1929)}]{Bloch1929}%
  \BibitemOpen
  \bibfield  {author} {\bibinfo {author} {\bibfnamefont {F.}~\bibnamefont
  {Bloch}},\ }\href@noop {} {\bibfield  {journal} {\bibinfo  {journal}
  {Zeitschrift f\"{u}r Physik}\ }\textbf {\bibinfo {volume} {52}},\ \bibinfo
  {pages} {555} (\bibinfo {year} {1929})}\BibitemShut {NoStop}%
\bibitem [{\citenamefont {Ashcroft}\ and\ \citenamefont
  {Mermin}(1976)}]{Ashcroft}%
  \BibitemOpen
  \bibfield  {author} {\bibinfo {author} {\bibfnamefont {N.~W.}\ \bibnamefont
  {Ashcroft}}\ and\ \bibinfo {author} {\bibfnamefont {D.~N.}\ \bibnamefont
  {Mermin}},\ }\href@noop {} {\emph {\bibinfo {title} {{Solid state
  physics}}}},\ \bibinfo {edition} {1st}\ ed.\ (\bibinfo  {publisher} {Thomson
  Learning},\ \bibinfo {address} {Toronto},\ \bibinfo {year}
  {1976})\BibitemShut {NoStop}%
\bibitem [{\citenamefont {Ashcroft}\ and\ \citenamefont
  {Lekner}(1966)}]{Ashcroft1966}%
  \BibitemOpen
  \bibfield  {author} {\bibinfo {author} {\bibfnamefont {N.~W.}\ \bibnamefont
  {Ashcroft}}\ and\ \bibinfo {author} {\bibfnamefont {J.}~\bibnamefont
  {Lekner}},\ }\href {\doibase 10.1103/PhysRev.145.83} {\bibfield  {journal}
  {\bibinfo  {journal} {Phys. Rev.}\ }\textbf {\bibinfo {volume} {145}},\
  \bibinfo {pages} {83} (\bibinfo {year} {1966})}\BibitemShut {NoStop}%
\bibitem [{\citenamefont {Neaton}\ and\ \citenamefont
  {Ashcroft}(1999)}]{Neaton1999}%
  \BibitemOpen
  \bibfield  {author} {\bibinfo {author} {\bibfnamefont {J.~B.}\ \bibnamefont
  {Neaton}}\ and\ \bibinfo {author} {\bibfnamefont {N.~W.}\ \bibnamefont
  {Ashcroft}},\ }\href {\doibase 10.1038/22067} {\bibfield  {journal} {\bibinfo
   {journal} {Nature}\ }\textbf {\bibinfo {volume} {400}},\ \bibinfo {pages}
  {141} (\bibinfo {year} {1999})}\BibitemShut {NoStop}%
\bibitem [{\citenamefont {Neaton}\ and\ \citenamefont
  {Ashcroft}(2001)}]{Neaton2001}%
  \BibitemOpen
  \bibfield  {author} {\bibinfo {author} {\bibfnamefont {J.~B.}\ \bibnamefont
  {Neaton}}\ and\ \bibinfo {author} {\bibfnamefont {N.~W.}\ \bibnamefont
  {Ashcroft}},\ }\href {\doibase 10.1103/PhysRevLett.86.2830} {\bibfield
  {journal} {\bibinfo  {journal} {Phys. Rev. Lett.}\ }\textbf {\bibinfo
  {volume} {86}},\ \bibinfo {pages} {2830} (\bibinfo {year}
  {2001})}\BibitemShut {NoStop}%
\bibitem [{\citenamefont {Rousseau}\ and\ \citenamefont
  {Ashcroft}(2008)}]{Rousseau2008}%
  \BibitemOpen
  \bibfield  {author} {\bibinfo {author} {\bibfnamefont {B.}~\bibnamefont
  {Rousseau}}\ and\ \bibinfo {author} {\bibfnamefont {N.~W.}\ \bibnamefont
  {Ashcroft}},\ }\href {\doibase 10.1103/PhysRevLett.101.046407} {\bibfield
  {journal} {\bibinfo  {journal} {Phys. Rev. Lett.}\ }\textbf {\bibinfo
  {volume} {101}},\ \bibinfo {pages} {046407} (\bibinfo {year}
  {2008})}\BibitemShut {NoStop}%
\bibitem [{\citenamefont {Landau}\ and\ \citenamefont
  {Lifshitz}(1980)}]{LandauStatMech}%
  \BibitemOpen
  \bibfield  {author} {\bibinfo {author} {\bibfnamefont {L.~D.}\ \bibnamefont
  {Landau}}\ and\ \bibinfo {author} {\bibfnamefont {E.}~\bibnamefont
  {Lifshitz}},\ }\href@noop {} {\emph {\bibinfo {title} {Statistical Physics
  I}}},\ Course of theoretical physics\ (\bibinfo  {publisher} {Elsevier},\
  \bibinfo {address} {Amsterdam},\ \bibinfo {year} {1980})\ \bibinfo {note} {p.
  168}\BibitemShut {NoStop}%
\bibitem [{\citenamefont {Plagemann}\ \emph {et~al.}(2012)\citenamefont
  {Plagemann}, \citenamefont {Sperling}, \citenamefont {Thiele}, \citenamefont
  {Desjarlais}, \citenamefont {Fortmann}, \citenamefont {D\"{o}ppner},
  \citenamefont {Lee}, \citenamefont {Glenzer},\ and\ \citenamefont
  {Redmer}}]{Plagemann2012}%
  \BibitemOpen
  \bibfield  {author} {\bibinfo {author} {\bibfnamefont {K.-U.}\ \bibnamefont
  {Plagemann}}, \bibinfo {author} {\bibfnamefont {P.}~\bibnamefont {Sperling}},
  \bibinfo {author} {\bibfnamefont {R.}~\bibnamefont {Thiele}}, \bibinfo
  {author} {\bibfnamefont {M.~P.}\ \bibnamefont {Desjarlais}}, \bibinfo
  {author} {\bibfnamefont {C.}~\bibnamefont {Fortmann}}, \bibinfo {author}
  {\bibfnamefont {T.}~\bibnamefont {D\"{o}ppner}}, \bibinfo {author}
  {\bibfnamefont {H.~J.}\ \bibnamefont {Lee}}, \bibinfo {author} {\bibfnamefont
  {S.~H.}\ \bibnamefont {Glenzer}}, \ and\ \bibinfo {author} {\bibfnamefont
  {R.}~\bibnamefont {Redmer}},\ }\href
  {http://stacks.iop.org/1367-2630/14/i=5/a=055020} {\bibfield  {journal}
  {\bibinfo  {journal} {New Journal of Physics}\ }\textbf {\bibinfo {volume}
  {14}},\ \bibinfo {pages} {055020} (\bibinfo {year} {2012})}\BibitemShut
  {NoStop}%
\bibitem [{\citenamefont {Vorberger}\ and\ \citenamefont
  {Gericke}(2013)}]{Vorberger2013}%
  \BibitemOpen
  \bibfield  {author} {\bibinfo {author} {\bibfnamefont {J.}~\bibnamefont
  {Vorberger}}\ and\ \bibinfo {author} {\bibfnamefont {D.}~\bibnamefont
  {Gericke}},\ }\href {\doibase http://dx.doi.org/10.1016/j.hedp.2012.12.009}
  {\bibfield  {journal} {\bibinfo  {journal} {High Energy Density Physics}\
  }\textbf {\bibinfo {volume} {9}},\ \bibinfo {pages} {178 } (\bibinfo {year}
  {2013})}\BibitemShut {NoStop}%
\bibitem [{\citenamefont {Itou}\ \emph {et~al.}(1998)\citenamefont {Itou},
  \citenamefont {Sakurai}, \citenamefont {Ohata}, \citenamefont {Bansil},
  \citenamefont {Kaprzyk}, \citenamefont {Tanaka}, \citenamefont {Kawata},\
  and\ \citenamefont {Shiotani}}]{Itou1998}%
  \BibitemOpen
  \bibfield  {author} {\bibinfo {author} {\bibfnamefont {M.}~\bibnamefont
  {Itou}}, \bibinfo {author} {\bibfnamefont {Y.}~\bibnamefont {Sakurai}},
  \bibinfo {author} {\bibfnamefont {T.}~\bibnamefont {Ohata}}, \bibinfo
  {author} {\bibfnamefont {A.}~\bibnamefont {Bansil}}, \bibinfo {author}
  {\bibfnamefont {S.}~\bibnamefont {Kaprzyk}}, \bibinfo {author} {\bibfnamefont
  {Y.}~\bibnamefont {Tanaka}}, \bibinfo {author} {\bibfnamefont
  {H.}~\bibnamefont {Kawata}}, \ and\ \bibinfo {author} {\bibfnamefont
  {N.}~\bibnamefont {Shiotani}},\ }\href {\doibase
  http://dx.doi.org/10.1016/S0022-3697(97)00146-7} {\bibfield  {journal}
  {\bibinfo  {journal} {Journal of Physics and Chemistry of Solids}\ }\textbf
  {\bibinfo {volume} {59}},\ \bibinfo {pages} {99 } (\bibinfo {year}
  {1998})}\BibitemShut {NoStop}%
\bibitem [{\citenamefont {Huotari}\ \emph {et~al.}(2000)\citenamefont
  {Huotari}, \citenamefont {H\"{a}m\"{a}l\"{a}inen}, \citenamefont {Manninen},
  \citenamefont {Kaprzyk}, \citenamefont {Bansil}, \citenamefont {Caliebe},
  \citenamefont {Buslaps}, \citenamefont {Honkim\"{a}ki},\ and\ \citenamefont
  {Suortti}}]{Huotari2000}%
  \BibitemOpen
  \bibfield  {author} {\bibinfo {author} {\bibfnamefont {S.}~\bibnamefont
  {Huotari}}, \bibinfo {author} {\bibfnamefont {K.}~\bibnamefont
  {H\"{a}m\"{a}l\"{a}inen}}, \bibinfo {author} {\bibfnamefont {S.}~\bibnamefont
  {Manninen}}, \bibinfo {author} {\bibfnamefont {S.}~\bibnamefont {Kaprzyk}},
  \bibinfo {author} {\bibfnamefont {A.}~\bibnamefont {Bansil}}, \bibinfo
  {author} {\bibfnamefont {W.}~\bibnamefont {Caliebe}}, \bibinfo {author}
  {\bibfnamefont {T.}~\bibnamefont {Buslaps}}, \bibinfo {author} {\bibfnamefont
  {V.}~\bibnamefont {Honkim\"{a}ki}}, \ and\ \bibinfo {author} {\bibfnamefont
  {P.}~\bibnamefont {Suortti}},\ }\href {\doibase 10.1103/PhysRevB.62.7956}
  {\bibfield  {journal} {\bibinfo  {journal} {Physical Review B}\ }\textbf
  {\bibinfo {volume} {62}},\ \bibinfo {pages} {7956} (\bibinfo {year}
  {2000})}\BibitemShut {NoStop}%
\bibitem [{\citenamefont {Huotari}\ \emph {et~al.}(2007)\citenamefont
  {Huotari}, \citenamefont {Sternemann}, \citenamefont {Volmer}, \citenamefont
  {Soininen}, \citenamefont {Monaco},\ and\ \citenamefont
  {Sch\"ulke}}]{Huotari2007}%
  \BibitemOpen
  \bibfield  {author} {\bibinfo {author} {\bibfnamefont {S.}~\bibnamefont
  {Huotari}}, \bibinfo {author} {\bibfnamefont {C.}~\bibnamefont {Sternemann}},
  \bibinfo {author} {\bibfnamefont {M.}~\bibnamefont {Volmer}}, \bibinfo
  {author} {\bibfnamefont {J.~A.}\ \bibnamefont {Soininen}}, \bibinfo {author}
  {\bibfnamefont {G.}~\bibnamefont {Monaco}}, \ and\ \bibinfo {author}
  {\bibfnamefont {W.}~\bibnamefont {Sch\"ulke}},\ }\href {\doibase
  10.1103/PhysRevB.76.235106} {\bibfield  {journal} {\bibinfo  {journal} {Phys.
  Rev. B}\ }\textbf {\bibinfo {volume} {76}},\ \bibinfo {pages} {235106}
  (\bibinfo {year} {2007})}\BibitemShut {NoStop}%
\bibitem [{\citenamefont {Huotari}\ \emph {et~al.}(2010)\citenamefont
  {Huotari}, \citenamefont {Soininen}, \citenamefont {Pylkk\"anen},
  \citenamefont {H\"am\"al\"ainen}, \citenamefont {Issolah}, \citenamefont
  {Titov}, \citenamefont {McMinis}, \citenamefont {Kim}, \citenamefont {Esler},
  \citenamefont {Ceperley}, \citenamefont {Holzmann},\ and\ \citenamefont
  {Olevano}}]{Huotari2010}%
  \BibitemOpen
  \bibfield  {author} {\bibinfo {author} {\bibfnamefont {S.}~\bibnamefont
  {Huotari}}, \bibinfo {author} {\bibfnamefont {J.~A.}\ \bibnamefont
  {Soininen}}, \bibinfo {author} {\bibfnamefont {T.}~\bibnamefont
  {Pylkk\"anen}}, \bibinfo {author} {\bibfnamefont {K.}~\bibnamefont
  {H\"am\"al\"ainen}}, \bibinfo {author} {\bibfnamefont {A.}~\bibnamefont
  {Issolah}}, \bibinfo {author} {\bibfnamefont {A.}~\bibnamefont {Titov}},
  \bibinfo {author} {\bibfnamefont {J.}~\bibnamefont {McMinis}}, \bibinfo
  {author} {\bibfnamefont {J.}~\bibnamefont {Kim}}, \bibinfo {author}
  {\bibfnamefont {K.}~\bibnamefont {Esler}}, \bibinfo {author} {\bibfnamefont
  {D.~M.}\ \bibnamefont {Ceperley}}, \bibinfo {author} {\bibfnamefont
  {M.}~\bibnamefont {Holzmann}}, \ and\ \bibinfo {author} {\bibfnamefont
  {V.}~\bibnamefont {Olevano}},\ }\href {\doibase
  10.1103/PhysRevLett.105.086403} {\bibfield  {journal} {\bibinfo  {journal}
  {Phys. Rev. Lett.}\ }\textbf {\bibinfo {volume} {105}},\ \bibinfo {pages}
  {086403} (\bibinfo {year} {2010})}\BibitemShut {NoStop}%
\bibitem [{\citenamefont {Herring}(1940)}]{Herring1940}%
  \BibitemOpen
  \bibfield  {author} {\bibinfo {author} {\bibfnamefont {C.}~\bibnamefont
  {Herring}},\ }\href {\doibase 10.1103/PhysRev.57.1169} {\bibfield  {journal}
  {\bibinfo  {journal} {Phys. Rev.}\ }\textbf {\bibinfo {volume} {57}},\
  \bibinfo {pages} {1169} (\bibinfo {year} {1940})}\BibitemShut {NoStop}%
\bibitem [{\citenamefont {Bl\"ochl}(1994)}]{Blochl1994}%
  \BibitemOpen
  \bibfield  {author} {\bibinfo {author} {\bibfnamefont {P.~E.}\ \bibnamefont
  {Bl\"ochl}},\ }\href {\doibase 10.1103/PhysRevB.50.17953} {\bibfield
  {journal} {\bibinfo  {journal} {Phys. Rev. B}\ }\textbf {\bibinfo {volume}
  {50}},\ \bibinfo {pages} {17953} (\bibinfo {year} {1994})}\BibitemShut
  {NoStop}%
\bibitem [{\citenamefont {Pandey}\ and\ \citenamefont
  {Lam}(1973)}]{Pandey1973}%
  \BibitemOpen
  \bibfield  {author} {\bibinfo {author} {\bibfnamefont {K.}~\bibnamefont
  {Pandey}}\ and\ \bibinfo {author} {\bibfnamefont {L.}~\bibnamefont {Lam}},\
  }\href {\doibase http://dx.doi.org/10.1016/0375-9601(73)90324-1} {\bibfield
  {journal} {\bibinfo  {journal} {Physics Letters A}\ }\textbf {\bibinfo
  {volume} {43}},\ \bibinfo {pages} {319 } (\bibinfo {year}
  {1973})}\BibitemShut {NoStop}%
\bibitem [{\citenamefont {Cooper}\ \emph {et~al.}(1974)\citenamefont {Cooper},
  \citenamefont {Pattison}, \citenamefont {Williams},\ and\ \citenamefont
  {Pandey}}]{Cooper1974}%
  \BibitemOpen
  \bibfield  {author} {\bibinfo {author} {\bibfnamefont {M.}~\bibnamefont
  {Cooper}}, \bibinfo {author} {\bibfnamefont {P.}~\bibnamefont {Pattison}},
  \bibinfo {author} {\bibfnamefont {B.}~\bibnamefont {Williams}}, \ and\
  \bibinfo {author} {\bibfnamefont {K.~C.}\ \bibnamefont {Pandey}},\
  }\href@noop {} {\bibfield  {journal} {\bibinfo  {journal} {Philosophical
  Magazine}\ }\textbf {\bibinfo {volume} {29}},\ \bibinfo {pages} {1237}
  (\bibinfo {year} {1974})}\BibitemShut {NoStop}%
\bibitem [{\citenamefont {Rennert}(1981)}]{Rennert1981}%
  \BibitemOpen
  \bibfield  {author} {\bibinfo {author} {\bibfnamefont {P.}~\bibnamefont
  {Rennert}},\ }\href@noop {} {\bibfield  {journal} {\bibinfo  {journal} {Phys.
  Stat. Sol. B}\ }\textbf {\bibinfo {volume} {105}},\ \bibinfo {pages} {567}
  (\bibinfo {year} {1981})}\BibitemShut {NoStop}%
\bibitem [{\citenamefont {Bellaiche}\ and\ \citenamefont
  {Kunc}(1997)}]{Bellaiche1997}%
  \BibitemOpen
  \bibfield  {author} {\bibinfo {author} {\bibfnamefont {L.}~\bibnamefont
  {Bellaiche}}\ and\ \bibinfo {author} {\bibfnamefont {K.}~\bibnamefont
  {Kunc}},\ }\href@noop {} {\bibfield  {journal} {\bibinfo  {journal} {Phys.
  Rev. B}\ }\textbf {\bibinfo {volume} {55}} (\bibinfo {year}
  {1997})}\BibitemShut {NoStop}%
\bibitem [{\citenamefont {Sch\"{u}lke}(2007)}]{Schuelke}%
  \BibitemOpen
  \bibfield  {author} {\bibinfo {author} {\bibfnamefont {W.}~\bibnamefont
  {Sch\"{u}lke}},\ }\href@noop {} {\emph {\bibinfo {title} {Electron Dynamics
  by Inelastic {X-Ray} Scattering}}},\ Oxford Series on Synchrotron Radiation\
  (\bibinfo  {publisher} {Oxford University Press},\ \bibinfo {address} {New
  York},\ \bibinfo {year} {2007})\BibitemShut {NoStop}%
\bibitem [{\citenamefont {Eisenberger}\ and\ \citenamefont
  {Platzman}(1970)}]{Eisenberger1970}%
  \BibitemOpen
  \bibfield  {author} {\bibinfo {author} {\bibfnamefont {P.}~\bibnamefont
  {Eisenberger}}\ and\ \bibinfo {author} {\bibfnamefont {P.~M.}\ \bibnamefont
  {Platzman}},\ }\href {\doibase 10.1103/PhysRevA.2.415} {\bibfield  {journal}
  {\bibinfo  {journal} {Physical Review A}\ }\textbf {\bibinfo {volume} {2}},\
  \bibinfo {pages} {415} (\bibinfo {year} {1970})}\BibitemShut {NoStop}%
\bibitem [{\citenamefont {Mattern}\ and\ \citenamefont
  {Seidler}(2013)}]{Mattern2013}%
  \BibitemOpen
  \bibfield  {author} {\bibinfo {author} {\bibfnamefont {B.~A.}\ \bibnamefont
  {Mattern}}\ and\ \bibinfo {author} {\bibfnamefont {G.~T.}\ \bibnamefont
  {Seidler}},\ }\href {\doibase 10.1063/1.4790659} {\bibfield  {journal}
  {\bibinfo  {journal} {Physics of Plasmas}\ }\textbf {\bibinfo {volume}
  {20}},\ \bibinfo {eid} {022706} (\bibinfo {year} {2013})}\BibitemShut
  {NoStop}%
\bibitem [{Sup()}]{Supplemental}%
  \BibitemOpen
  \href@noop {} {}\bibinfo {note} {See Supplemental Information}\BibitemShut
  {NoStop}%
\bibitem [{\citenamefont {Kas}\ \emph {et~al.}(2013)\citenamefont {Kas},
  \citenamefont {Rehr},\ and\ \citenamefont {Reining}}]{Kas2013}%
  \BibitemOpen
  \bibfield  {author} {\bibinfo {author} {\bibfnamefont {J.~J.}\ \bibnamefont
  {Kas}}, \bibinfo {author} {\bibfnamefont {J.~J.}\ \bibnamefont {Rehr}}, \
  and\ \bibinfo {author} {\bibfnamefont {L.}~\bibnamefont {Reining}},\
  }\href@noop {} {\bibfield  {journal} {\bibinfo  {journal} {\textit{To be
  submitted} Phys. Rev. Lett.}\ } (\bibinfo {year} {2013})}\BibitemShut
  {NoStop}%
\bibitem [{\citenamefont {Benedict}\ \emph {et~al.}(2009)\citenamefont
  {Benedict}, \citenamefont {Ogitsu}, \citenamefont {Trave}, \citenamefont
  {Wu}, \citenamefont {Sterne},\ and\ \citenamefont
  {Schwegler}}]{Benedict2009}%
  \BibitemOpen
  \bibfield  {author} {\bibinfo {author} {\bibfnamefont {L.~X.}\ \bibnamefont
  {Benedict}}, \bibinfo {author} {\bibfnamefont {T.}~\bibnamefont {Ogitsu}},
  \bibinfo {author} {\bibfnamefont {A.}~\bibnamefont {Trave}}, \bibinfo
  {author} {\bibfnamefont {C.~J.}\ \bibnamefont {Wu}}, \bibinfo {author}
  {\bibfnamefont {P.~A.}\ \bibnamefont {Sterne}}, \ and\ \bibinfo {author}
  {\bibfnamefont {E.}~\bibnamefont {Schwegler}},\ }\href {\doibase
  10.1103/PhysRevB.79.064106} {\bibfield  {journal} {\bibinfo  {journal} {Phys.
  Rev. B}\ }\textbf {\bibinfo {volume} {79}},\ \bibinfo {pages} {064106}
  (\bibinfo {year} {2009})}\BibitemShut {NoStop}%
\bibitem [{\citenamefont {Mattern}\ \emph {et~al.}(2012)\citenamefont
  {Mattern}, \citenamefont {Seidler}, \citenamefont {Kas}, \citenamefont
  {Pacold},\ and\ \citenamefont {Rehr}}]{Mattern2012}%
  \BibitemOpen
  \bibfield  {author} {\bibinfo {author} {\bibfnamefont {B.~A.}\ \bibnamefont
  {Mattern}}, \bibinfo {author} {\bibfnamefont {G.~T.}\ \bibnamefont
  {Seidler}}, \bibinfo {author} {\bibfnamefont {J.~J.}\ \bibnamefont {Kas}},
  \bibinfo {author} {\bibfnamefont {J.~I.}\ \bibnamefont {Pacold}}, \ and\
  \bibinfo {author} {\bibfnamefont {J.~J.}\ \bibnamefont {Rehr}},\ }\href
  {\doibase 10.1103/PhysRevB.85.115135} {\bibfield  {journal} {\bibinfo
  {journal} {Phys. Rev. B}\ }\textbf {\bibinfo {volume} {85}},\ \bibinfo
  {pages} {115135} (\bibinfo {year} {2012})}\BibitemShut {NoStop}%
\bibitem [{\citenamefont {Rehr}\ \emph {et~al.}(2009)\citenamefont {Rehr},
  \citenamefont {Kas}, \citenamefont {Prange}, \citenamefont {Sorini},
  \citenamefont {Takimoto},\ and\ \citenamefont {Vila}}]{Rehr2009}%
  \BibitemOpen
  \bibfield  {author} {\bibinfo {author} {\bibfnamefont {J.~J.}\ \bibnamefont
  {Rehr}}, \bibinfo {author} {\bibfnamefont {J.~J.}\ \bibnamefont {Kas}},
  \bibinfo {author} {\bibfnamefont {M.~P.}\ \bibnamefont {Prange}}, \bibinfo
  {author} {\bibfnamefont {A.~P.}\ \bibnamefont {Sorini}}, \bibinfo {author}
  {\bibfnamefont {Y.}~\bibnamefont {Takimoto}}, \ and\ \bibinfo {author}
  {\bibfnamefont {F.}~\bibnamefont {Vila}},\ }\href {\doibase
  http://dx.doi.org/10.1016/j.crhy.2008.08.004} {\bibfield  {journal} {\bibinfo
   {journal} {Comptes Rendus Physique}\ }\textbf {\bibinfo {volume} {10}},\
  \bibinfo {pages} {548 } (\bibinfo {year} {2009})}\BibitemShut {NoStop}%
\bibitem [{\citenamefont {Rehr}\ and\ \citenamefont
  {Albers}(2000)}]{RehrAlbers2000}%
  \BibitemOpen
  \bibfield  {author} {\bibinfo {author} {\bibfnamefont {J.~J.}\ \bibnamefont
  {Rehr}}\ and\ \bibinfo {author} {\bibfnamefont {R.~C.}\ \bibnamefont
  {Albers}},\ }\href {\doibase 10.1103/RevModPhys.72.621} {\bibfield  {journal}
  {\bibinfo  {journal} {Rev. Mod. Phys.}\ }\textbf {\bibinfo {volume} {72}},\
  \bibinfo {pages} {621} (\bibinfo {year} {2000})}\BibitemShut {NoStop}%
\bibitem [{\citenamefont {Epstein}(1973)}]{Epstein1973}%
  \BibitemOpen
  \bibfield  {author} {\bibinfo {author} {\bibfnamefont {I.~R.}\ \bibnamefont
  {Epstein}},\ }\href {\doibase 10.1103/PhysRevA.8.160} {\bibfield  {journal}
  {\bibinfo  {journal} {Phys. Rev. A}\ }\textbf {\bibinfo {volume} {8}},\
  \bibinfo {pages} {160} (\bibinfo {year} {1973})}\BibitemShut {NoStop}%
\bibitem [{\citenamefont {Perrot}\ and\ \citenamefont
  {Dharma-wardana}(1984)}]{Perrot1984}%
  \BibitemOpen
  \bibfield  {author} {\bibinfo {author} {\bibfnamefont {F.~m.~c.}\
  \bibnamefont {Perrot}}\ and\ \bibinfo {author} {\bibfnamefont {M.~W.~C.}\
  \bibnamefont {Dharma-wardana}},\ }\href {\doibase 10.1103/PhysRevA.30.2619}
  {\bibfield  {journal} {\bibinfo  {journal} {Phys. Rev. A}\ }\textbf {\bibinfo
  {volume} {30}},\ \bibinfo {pages} {2619} (\bibinfo {year}
  {1984})}\BibitemShut {NoStop}%
\bibitem [{\citenamefont {Mattern}(2013)}]{MatternThesis}%
  \BibitemOpen
  \bibfield  {author} {\bibinfo {author} {\bibfnamefont {B.~A.}\ \bibnamefont
  {Mattern}},\ }\emph {\bibinfo {title} {Compton Scattering and Warm Dense
  Matter Thermometry}},\ \href@noop {} {Ph.D. thesis},\ \bibinfo  {school}
  {University of Washington, Seattle} (\bibinfo {year} {2013})\BibitemShut
  {NoStop}%
\end{thebibliography}
\end{document}